\documentclass[a4paper,11pt]{article}
\usepackage{pos,bm}
\usepackage{multicol,hyperref}

\newcommand{\beq}{\begin{equation}}
\newcommand{\eeq}{\end{equation}}
\newcommand{\beqs}{\begin{eqnarray}}
\newcommand{\eeqs}{\end{eqnarray}}
\newcommand{\me}[3]{\langle #1\vert\ #2\ \vert #3\rangle}

\title{Lattice QCD studies of the $\Delta$ baryon resonance and the $K_0^\ast(700)$ and 
       $a_0(980)$ meson resonances: the role of exotic operators in determining the 
       finite-volume spectrum}
\ShortTitle{$\Delta$, $K_0^\ast(700)$, and $a_0(980)$}

\author[a]{John Bulava}
\author[b]{Danny Darvish}
\author[c]{Andrew D. Hanlon}
\author[d]{Ben H\"{o}rz}
\author[b]{Colin Morningstar}
\author[e]{Amy Nicholson}
\author[f]{Fernando Romero-L\'{o}pez}
\author*[b]{Sarah Skinner}
\author[g,h]{Pavlos Vranas}
\author[h]{Andr\'{e} Walker-Loud}

\affiliation[a]{Fakult\"{a}t f\"{u}r Physik und Astronomie, Institut f\"{u}r Theoretische Physik II, Ruhr-Universit\"{a}t Bochum, 44780 Bochum, Germany}

\affiliation[b]{Department of Physics, Carnegie Mellon University, 
Pittsburgh, Pennsylvania 15213, USA}

\affiliation[c]{Physics Department, Brookhaven National Laboratory,
Upton, New York 11973, USA}

\affiliation[d]{Intel Deutschland GmbH, Dornacher Str. 1, 85622 Feldkirchen, Germany}

\affiliation[e]{Department of Physics and Astronomy, University of North Carolina, 
  Chapel Hill, NC 27516, USA}

\affiliation[f]{Center for Theoretical Physics, Massachusetts Institute of Technology, 
  Cambridge, MA 02139, USA}

\affiliation[g]{Physics Division, Lawrence Livermore National Laboratory, 94550, 
  Livermore, CA, USA}

\affiliation[h]{Nuclear Science Division, Lawrence Berkeley National Laboratory, 
Berkeley, California 94720, USA}

\emailAdd{sarahski@andrew.cmu.edu}
\emailAdd{cmorning@andrew.cmu.edu}

\abstract{Studies of the $\Delta$ baryon resonance and the $K_0^\ast(700)$ and $a_0(980)$ meson 
  resonances using $N_f=2+1$ lattice QCD for pion masses near 200 MeV are presented.
  The $s$-wave scattering lengths for both the $I=1/2$ $N \pi$ and $I=3/2$ $N \pi$ channels 
  and properties of the $\Delta$ resonance are identified from the finite-volume energy levels 
  of the lattice simulation. The importance of a three-quark $\Delta$-operator in the $N\pi$ 
  system and tetraquark operators in the mesonic systems is investigated.}

\FullConference{The 40th International Symposium on Lattice Field Theory (Lattice 2023)\\
July 31st - August 4th, 2023\\
Fermi National Accelerator Laboratory\\}

\begin{document}
\maketitle

\section{Overview}

This talk presents our recent results\cite{Bulava:2022vpq} for $I=\frac{1}{2}, \frac{3}{2}$ $N\pi$ 
scattering amplitudes including the $\Delta(1232)$ resonance using the D200 ensemble from the 
Coordinated Lattice Simulations (CLS) consortium with $m_\pi=200$~MeV and $N_f=2+1$ dynamical 
fermions, as well as preliminary results from our ongoing study\cite{Darvish:2019oie} of the 
$K_0^\ast(700)$ and $a_0(980)$ meson resonances.  The importance of including a local three-quark
$\Delta$ operator in our $N\pi$ studies and a tetraquark operator in studies of the above mesonic 
systems is demonstrated and highlighted.

To obtain two-hadron to two-hadron scattering amplitudes and resonance information in lattice 
QCD, the first step is to compute the finite-volume (FV) energy spectra in the relevant 
channels of interest involving the scattered hadrons. Once the FV spectra are obtained, the 
elements of the infinite-volume two-hadron to two-hadron scattering $K$-matrix must be 
appropriately parametrized, then best fit values of the parameters are obtained using 
the L\"uscher quantization 
condition\cite{Luscher:1990ux,Rummukainen:1995vs,Kim:2005gf,Briceno:2014oea}.  
Details of this
entire procedure can be found in Ref.~\cite{Bulava:2022vpq}.  Our implementation of the 
L\"uscher method uses the ``box matrix'' $B$ introduced in Ref.~\cite{Morningstar:2017spu}, 
along with the scattering $K$-matrix. 
In this talk, we focus on the first step, extracting the FV spectra, highlighting the 
importance of using judiciously chosen interpolating operators.
 
\section{Finite-volume spectra}

In lattice QCD, FV energy spectra are retrieved from matrices of temporal correlations 
$C_{ij}(t)= \langle 0 | \mathcal{O}_i(t+t_0) \overline{\mathcal{O}}_j(t_0) | 0 \rangle$,
where $\mathcal{O}_i(t)$ denotes the $i$-th interpolating operator constructed from quark 
and gluon fields to create single- and multi-hadron states having appropriate transformation 
properties. In finite volume, stationary-state energies are discrete, so by inserting a 
complete set of states, the correlators can be expressed in terms of the energies using   
   \begin{equation}
   C_{ij}(t) = \sum_{n=0}^\infty Z_i^{(n)} Z_j^{(n)\ast}\ e^{-E_n t},
   \qquad\quad Z_j^{(n)}=  \me{0}{O_j}{n},
   \end{equation}
ignoring negligible effects from the temporal boundary. 
Given the large number of complex-valued overlap factors in the above equation, it is not 
practical to do simultaneous fits to the entire $C(t)$ matrix, so instead, we carry out 
separate fits to each of the diagonal elements of the matrix $\widetilde{C}(t)$ 
obtained using a single-pivot rotation\cite{Fox:1981xz,Michael:1982gb,Luscher:1990ck}
   \begin{equation}
   \widetilde{C}(t) = U^\dagger\ C(\tau_0)^{-1/2}\ C(t)\ C(\tau_0)^{-1/2}\ U,
   \end{equation}
where the  columns of $U$ are the eigenvectors of
   $C(\tau_0)^{-1/2}\,C(\tau_D)\,C(\tau_0)^{-1/2}$, determined
by solving a generalized eigenvector problem (GEVP).
We choose $\tau_0$ and $\tau_D$ large enough so that $\widetilde{C}(t)$ 
remains diagonal for $t>\tau_D$ and such that the extracted energies are
insensitive to increases in these parameters.
Typically, two-exponential fits to the diagonal elements $\widetilde{C}_{\alpha\alpha}(t)$
yield the energies $E_\alpha$ and overlaps $Z_j^{(n)}$.
However, a variety of
other fits are often employed as cross-checks, as detailed in, for example,
Ref.~\cite{Bulava:2022vpq}.

Since stochastic estimates of $C_{ij}(t)$ are obtained using the Monte Carlo method and 
the signal-to-noise ratios of such estimates degrade quickly with $t$, it is very important 
to use judiciously constructed operators $O_j(t)$ to maximize $Z$ factors for the states 
of interest and minimize those for unwanted higher-lying states in order to reliably reveal 
the low-lying energies.
Our operator construction is detailed in Refs.\cite{PhysRevD.72.094506,PhysRevD.88.014511}.
Individual hadron operators are constructed using basic building blocks which
are covariantly-displaced LapH-smeared quark fields\cite{Morningstar:2011ka} with 
stout link smearing\cite{PhysRevD.69.054501}.
Our hadron operator construction is very efficient and generalizes to
three or more hadrons.  Note that to speed up our computations to achieve the
statistics needed for extracting the low-lying energies required for our 
meson-baryon scattering studies, we have not included any single hadron 
operators with quarks that are displaced from one another in the $\Delta$ study.
Including multi-hadron operators in our correlation matrices requires the
use of time-slice to time-slice quark propagators.  To make the calculations
feasible, we resort to employing stochastic estimates of such quark propagators.
The stochastic LapH method\cite{Morningstar:2011ka} is used.

If the $N$ lowest-lying energies in a particular symmetry channel are needed, a correlation 
matrix using at least $N$ interpolating operators must be evaluated.  To improve the energy 
extractions, more than $N$ operators are usually required to help remove contamination from 
levels above the lowest $N$ states.  A crucial point, however, is that in practice, for each 
of the $N$ lowest-lying energies, an operator should be present which produces a state having 
significant overlap with the eigenstate associated with that energy.
Without such an operator set, the limited temporal range 
of the correlations which can be reliably estimated due to the limited statistics 
possible with current Monte Carlo methods can lead to missed energy levels.

For each symmetry channel of interest specified by a total momentum and an irreducible
representation (irrep) of its little group, a basic set of operators to use should include 
operators that contain the incoming particles as well as operators that contain the outgoing 
particles of the scattering process in all of the allowed individual momenta and in 
spin/orbital combinations that transform according
to the irrep of the channel.  Determining if \textit{other} operators are also needed and 
what those operators should be is an important issue.

\begin{table}[t]
\begin{center}
\caption{Lattice dimensions, masses of the pion, kaon, and nucleon, decay constants
 of the pion and kaon, and number of configurations for the $N\pi$ scattering analysis (taken from 
 Ref.~\cite{Bulava:2022vpq})
\label{tab:comp_deets_mb}}
\footnotesize
\begin{tabular}{c@{\hskip 12pt}c@{\hskip 12pt}c@{\hskip 12pt}c@{\hskip 12pt}c@{\hskip 12pt}c@{\hskip 12pt}c@{\hskip 12pt}c@{\hskip 12pt}}
 $a [\textup{fm}]$ & $(L/a)^3 \times T/a$ & $N_{\rm meas}$ & $am_\pi$ &  $am_{\rm K}$ & $af_{\pi}$ & $af_{\rm K}$ & $am_{\rm N}$ \\ \hline
     0.0633(4)(6) & $64^3 \times 128$ & 2000 & 0.06617(33) & 0.15644(16) & 0.04233(16) & 0.04928(21) &  0.3148(23)
\end{tabular}
\end{center}
\end{table}

\section{$\mathbf{\Delta}$ resonance}
\label{sec:delta_analysis}

\begin{figure}
    \centering
    \includegraphics[width=0.49\linewidth]{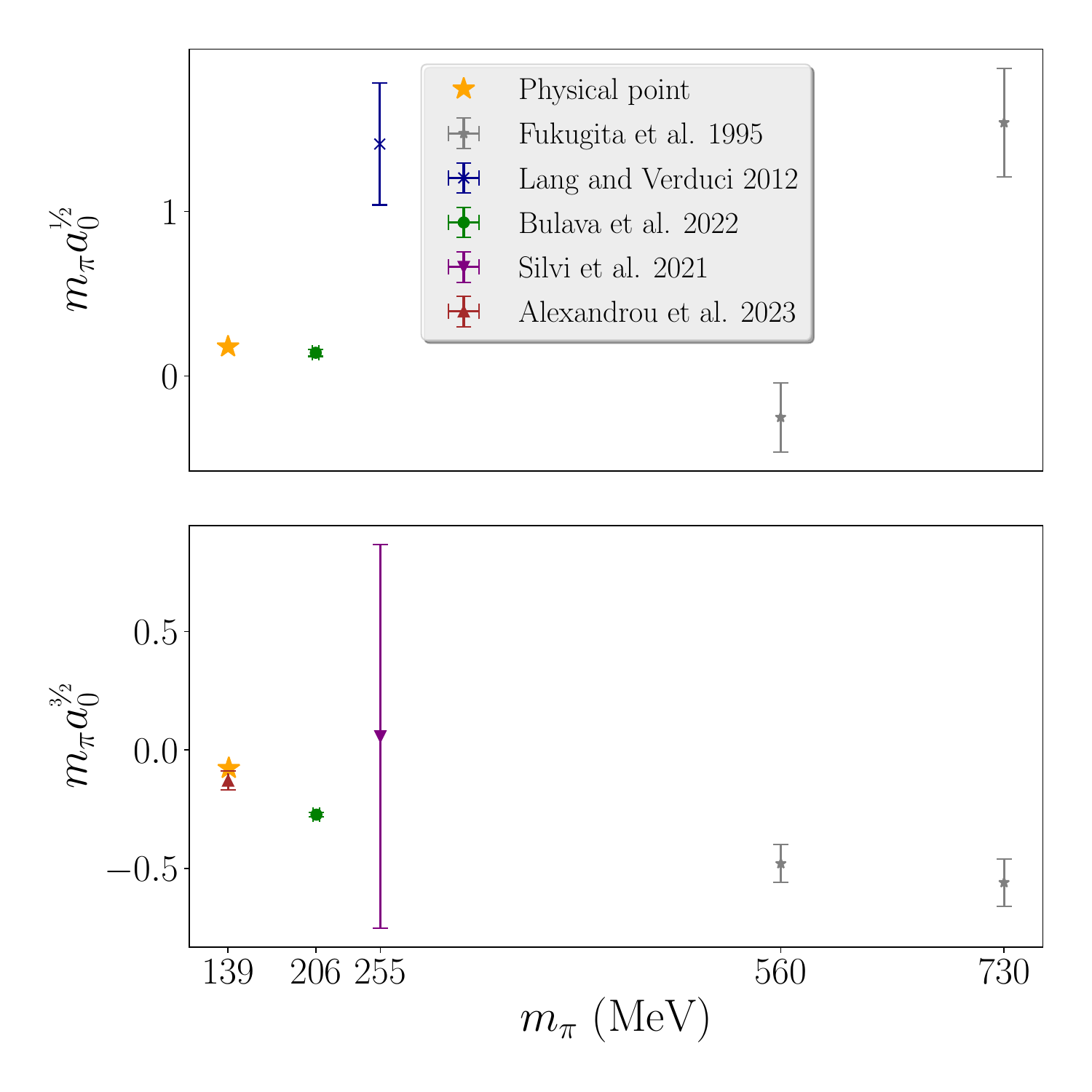}
    \hfill
    \includegraphics[width=0.49\linewidth]{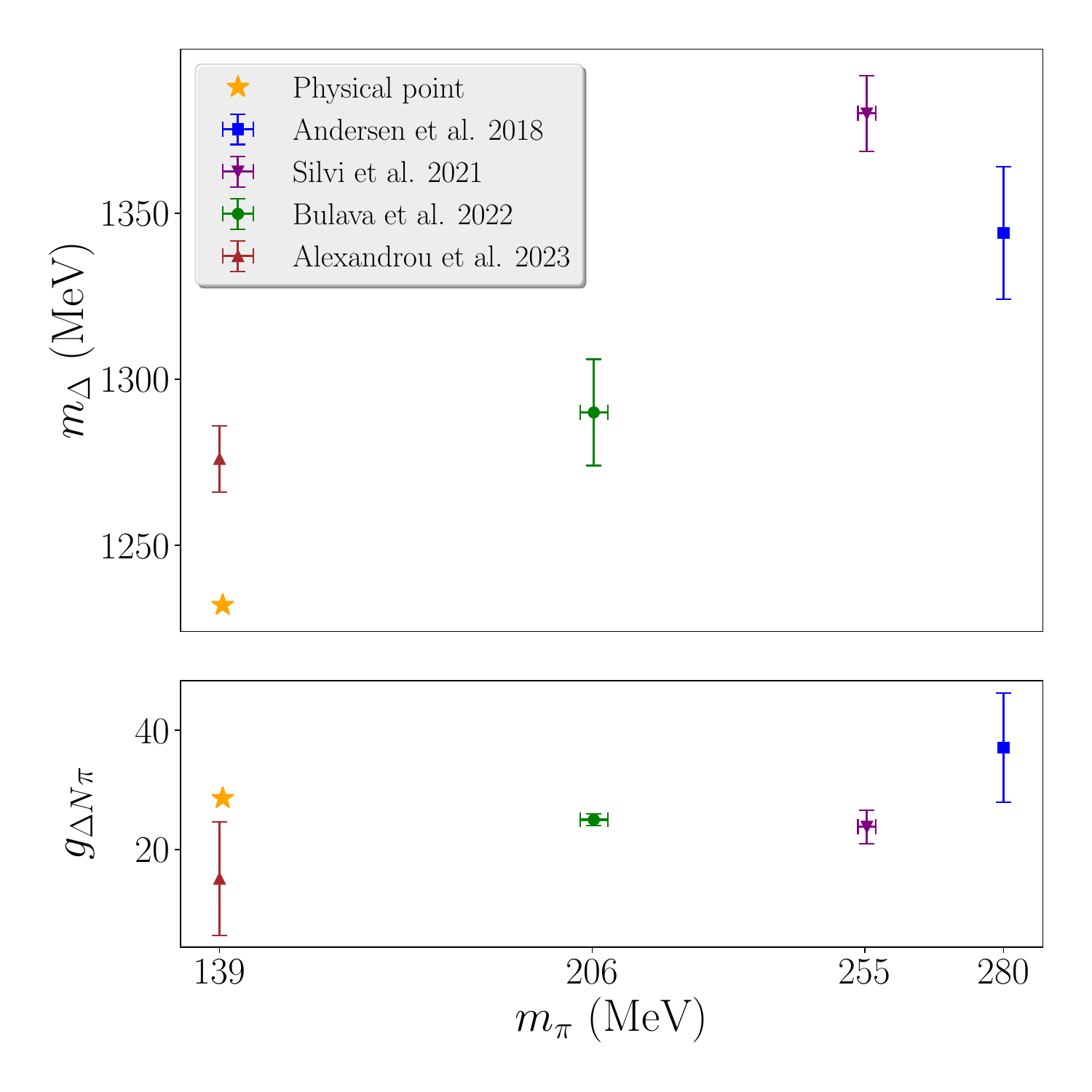}
    \caption{Comparison of our results (green) with other recent works for the
(left) $I=1/2$ and $I=3/2$ scattering lengths, $a_0^{1/2}$ and $a_0^{3/2}$ and
 the (right) Breit-Wigner mass, $m_\Delta$ and the coupling $g_{\Delta N\pi}$ from 
 leading-order effective field theory.
 \label{fig:npilandscape}}
\end{figure}

Our study of $N\pi$ scattering has appeared in Ref~\cite{Bulava:2022vpq}, and
the details of the CLS D200 Monte Carlo ensemble are given in Table~\ref{tab:comp_deets_mb}.
A comparison of our results with other recent works for scattering lengths and
the $\Delta$ resonance mass and its width parameter is shown in Fig.~\ref{fig:npilandscape}.

\begin{figure}
 \centering
 \includegraphics[width=0.32\linewidth]{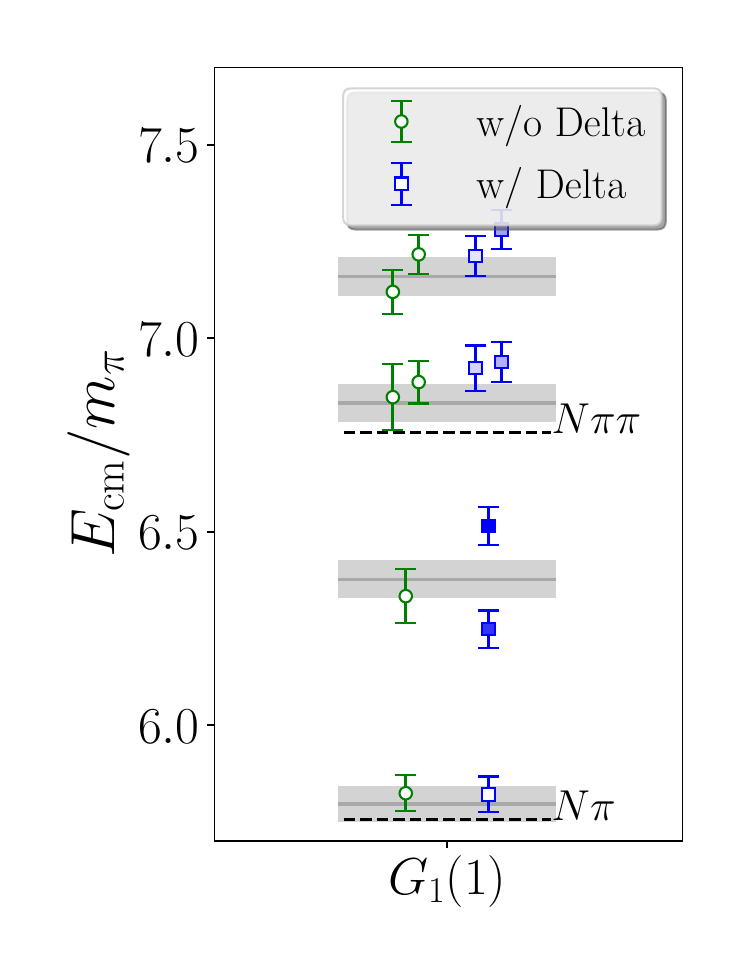}
 \includegraphics[width=0.32\linewidth]{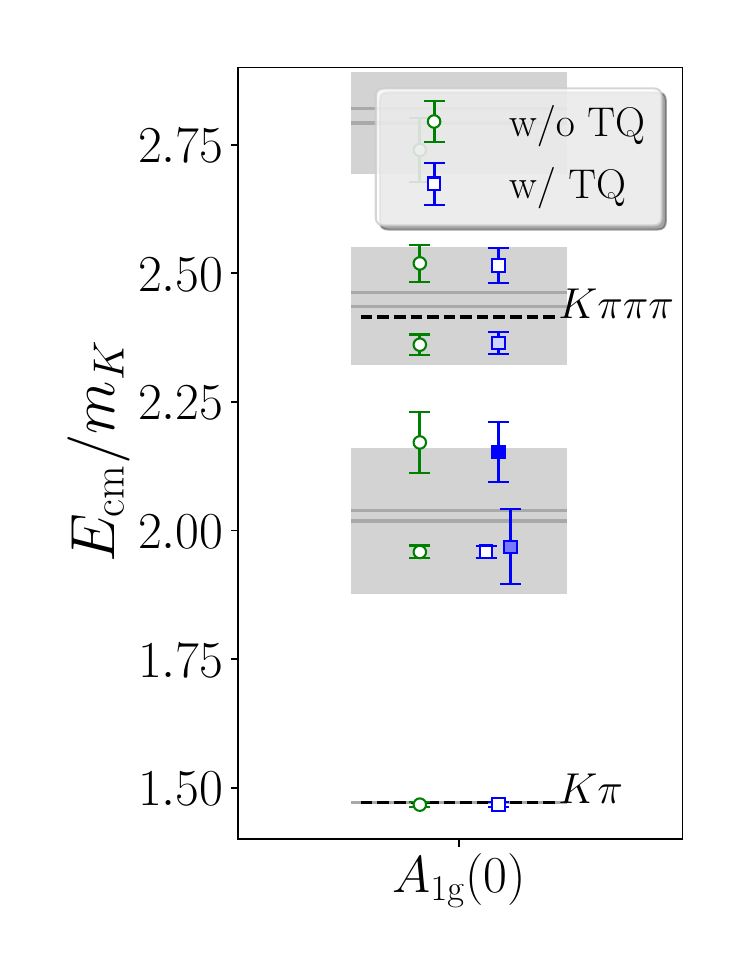}
 \includegraphics[width=0.32\linewidth]{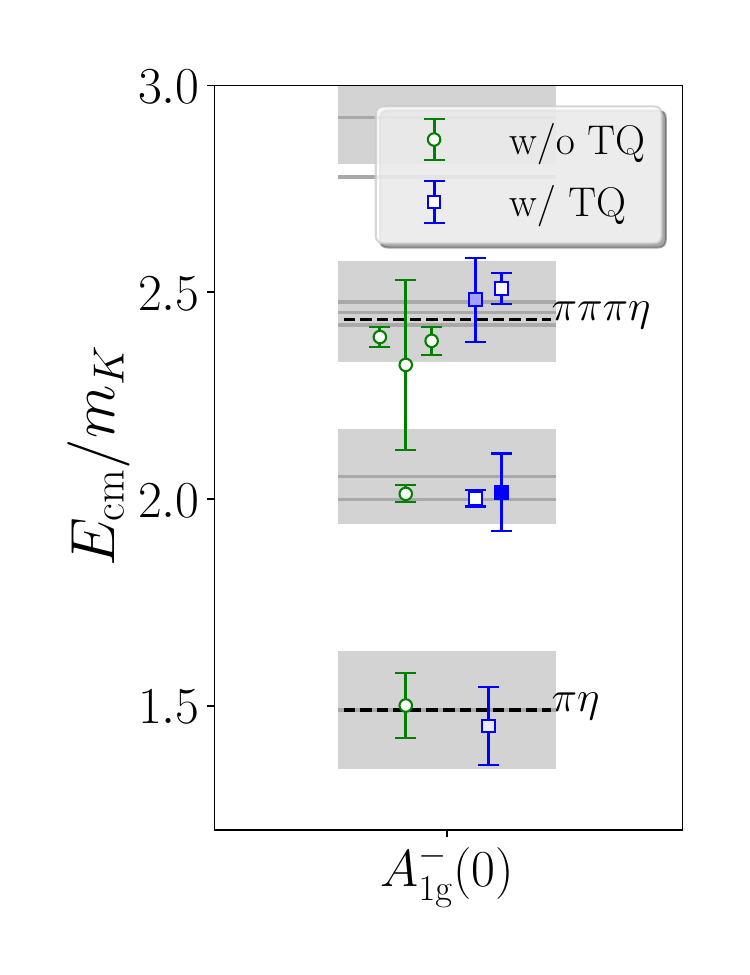}
 \caption{Effects on energy spectrum extractions when missing an important interpolating operator.
  (Left) Isoquartet $N\pi$ channel $G_1(1)$ irrep with and without the local three-quark $\Delta$ 
  operator. 
  (Center) Isodoublet strange channel $A_{1g}(0)$ irrep with and without the tetraquark operator.
  (Right) Isotriplet nonstrange channel $A_{1g}^-(0)$ irrep with and without the tetraquark
  operator. The degree of fill of the plot symbols represents the relative magnitude of the 
corresponding exotic ($\Delta$ or TQ) operator overlaps with those levels.
  \label{fig:spectrums}}
\end{figure}

\begin{figure}
 \centering
  \fcolorbox{green}{white}{\includegraphics[trim={1.3cm 0 0 0},clip,width=0.3\linewidth]{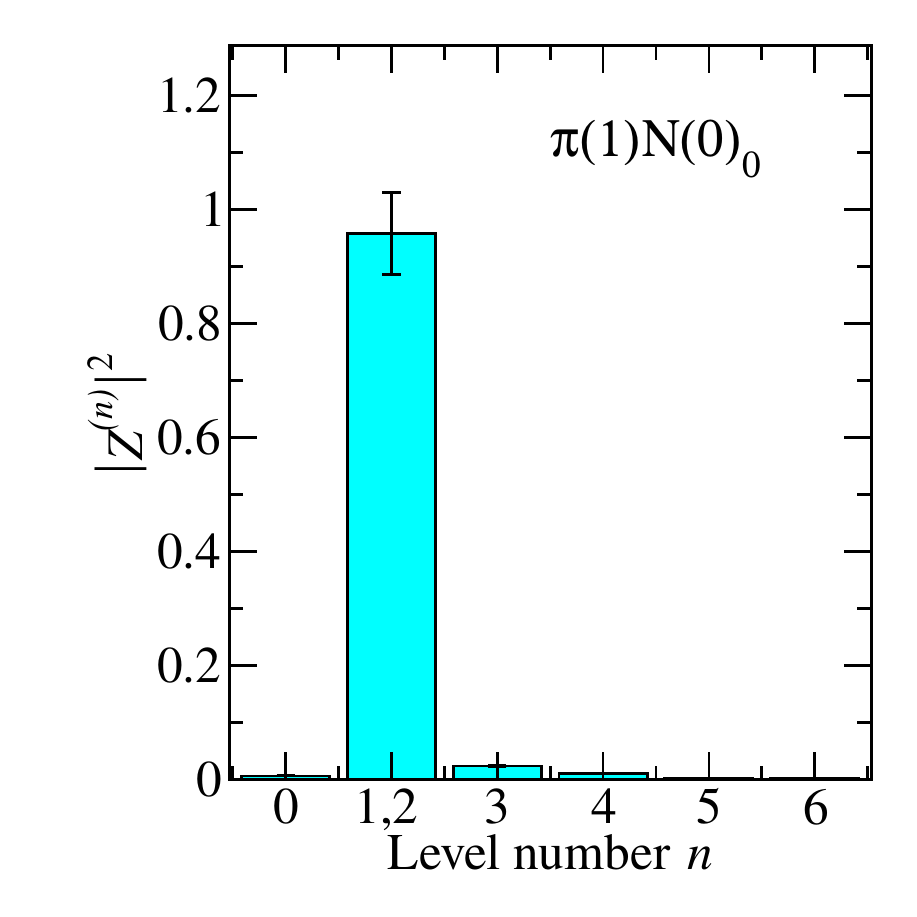}}
  \hfill
  \fcolorbox{blue}{white}{\includegraphics[trim={0.5cm 0 0 0},clip,width=0.3\linewidth]{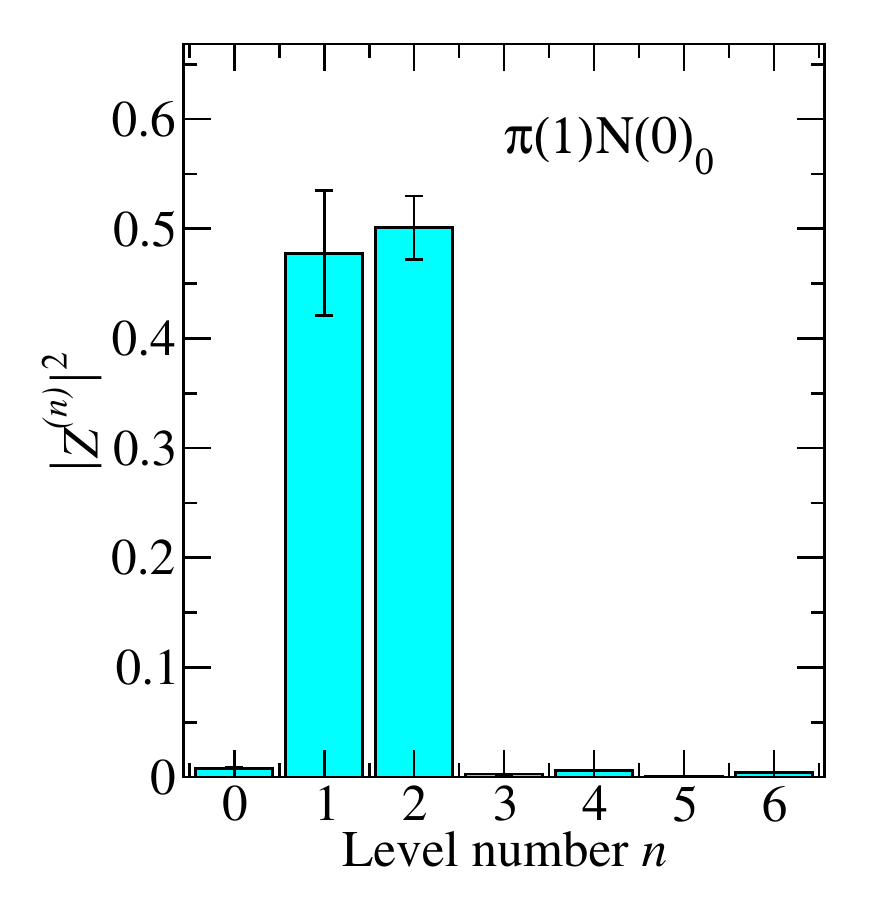}}
  \hfill
  \fcolorbox{blue}{white}{\includegraphics[trim={0.5cm 0 0 0},clip,width=0.3\linewidth]{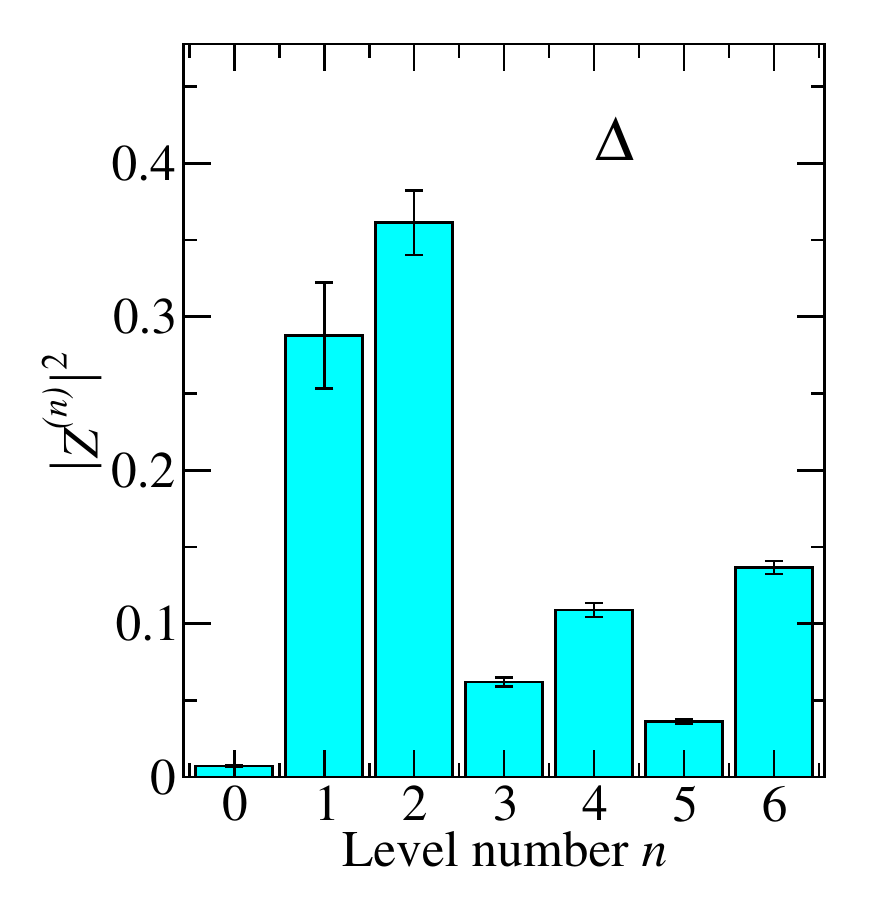}}
  \caption{Isoquartet $\pi(1)N(0)_0$ operator overlap factors without (left) and with (center) 
  the $\Delta$ 
  operator included in the correlation matrix. (Right) $\Delta$ operator overlap factors.
  Magnitudes are normalized within each plot and level number ordering is based on increasing
  final energy fit values of the spectrum including the $\Delta$ operator. 
  \label{fig:delta_ops}}
\end{figure}

What was not included in Ref~\cite{Bulava:2022vpq} was a deeper study of the effect of the 
local three-quark $\Delta$ operator on our ability to extract the FV spectrum. 
The left panel of Fig.~\ref{fig:spectrums} reveals that the three-quark $\Delta$
operator is crucial for reliably determining the FV spectrum.  This plot shows
the spectrum extracted with all possible $N\pi$ operators and with and without including 
the $\Delta$ operator.  Excluding this operator causes us to miss an energy level, as 
well as making our determinations of nearby energies less reliable. To separate out
levels 1 and 2, both the $\pi(1)N(0)_0$ and the $\Delta$ operator are needed.  The integers in the parentheses indicate $\bm{d}^2$ of the particle, where the momentum of the particle is $\bm{p}=2\pi\bm{d}/L$, and the
subscript 0 in the $N\pi$ operator denotes one particular Clebsch-Gordan combination of 
the baryon-meson fields. The middle
and right plots of Fig.~\ref{fig:delta_ops} show the overlaps onto levels 1 and 2
of the $\pi(1)N(0)_0$ operator and the $\Delta$ operator.  Without the $\Delta$
operator, levels 1 and 2 cannot be separated out since their energies are so close
and overlaps of these two levels with the states created by all other operators are 
negligible. The overlaps onto these two states 
get combined into a single overlap factor for the $\pi(1)N(0)_0$ operator, as shown in the 
left plot of Fig.~\ref{fig:delta_ops}.

\section{$\mathbf{K_0^*, a_0}$ study}

We have also studied the isodoublet strange $A_{1g}(0)$ channel, which should contain the
$K_0^\ast(700)$ resonance, as well as the isotriplet nonstrange $A_{1g}^-(0)$ channel,
which should contain the $a_0(980)$ resonance.  In the irrep notation, the zero in
parentheses indicates a channel of zero total momentum, the subscript $g$ refers to
even parity, and the superscript $-$ indicates odd $G$-parity.  The ensemble details
for our study are given in Tab.~\ref{tab:comp_deets_mm}.  Note that to achieve
reliable quantitative studies of the $K_0^\ast(700)$ and $a_0(980)$ resonances, a
variety of channels having different total momenta is needed to map out the resonance
features.  Here, our goal is more qualitative to focus on the importance of so-called
tetraquark operators for extracting the FV spectra.  Hence, we limit our attention
to channels having only zero total momentum.

\begin{table}[t]
\caption{Details of the ensemble used for our meson-meson analysis. $a_s$ and $a_t$ are the spacial 
and temporal lattice spacing, $\xi=a_s/a_t$ is the anisotropy, $L/a_s$ and $T/a_t$ represent 
total lengths of the lattice in the spacial and temporal directions, $N_{\rm meas}$ is the number 
of configurations sampled, and $a_tm_\pi$ and $a_tm_K$ are the pion and kaon masses in lattice
units. \label{tab:comp_deets_mm}}
\begin{center}
\footnotesize
\begin{tabular}{c@{\hskip 12pt}c@{sa\hskip 12pt}c@{\hskip 12pt}c@{\hskip 12pt}c@{\hskip 12pt}c@{\hskip 12pt}c@{\hskip 12pt}c@{\hskip 12pt}}
 $a_t [\textup{fm}]$ & $\xi$ & $(L/a_s)^3 \times T/a_t$ & $N_{\rm meas}$ & $a_tm_\pi$ &  $a_tm_{\rm K}$  \\ \hline
     0.033357(59) & 3.451(11) & $32^3 \times 256$ & 412 & 0.06617(33) & 0.15644(16) 
\end{tabular}
\end{center}
\end{table}

In the isodoublet channel, we include three quark-antiquark extended operators 
in which the quark is displaced in some way from the antiquark.  Four $K\pi$ operators
are used, and two $K\eta$ and two $K\phi$ operators are also included.  Here, $\eta$ 
refers to an isosinglet operator having flavor structure $\overline{u}u+\overline{d}d$, and $\phi$ 
refers to an isosinglet operator having flavor structure $\overline{s}s$.  In the
isotriplet channel, two extended quark-antiquark operators are used, four $\overline{K}K$
operators are used, and three $\pi\eta$ and three $\pi\phi$ operators are also included.

Some past works have suggested that these resonances might require tetraquark operators
to be investigated reliably\cite{Prelovsek:2010kg,Alexandrou:2017itd}.  To test this, we 
designed and implemented a large variety of tetraquark operators.  Several hundred tetraquark 
operators of different spatial and orbital structure, as well as different flavor structure, 
were studied.  All of our tetraquark operators are constructed out of two quarks and two 
antiquarks.  Although the color structure is very similar to that of a meson-meson operator, 
the individual color-contracted quark-antiquark pairs in a tetraquark are not each formed with 
a separate individual well-defined momentum nor does each pair transform irreducibly under 
any symmetry except color. Only the full combination of the two quark-antiquark pairs
is formed with well defined momentum and spin/orbital transformation properties.

As described in Refs.~\cite{Darvish:2019oie,DarvishThesis},
in each of the isodoublet and the isotriplet channels, a spectrum was first extracted
using all of the meson and meson-meson operators, but excluding any tetraquark operators.
Secondly, low-statistics spectra were extracted using all of the meson and meson-meson operators
while including one tetraquark operator.  This was done for all of the hundreds of
tetraquark operators devised using 25 configurations.  Most of the tetraquark operators
did not result in an additional low-lying energy level, but in both the isodoublet  and
isotriplet cases, about a dozen or so tetraquark operators did yield an additional
level.  Thirdly, choosing only from the subset of tetraquark operators that did create 
an additional level, we extracted spectra for operator sets which included two 
tetraquark operators, again using only 25 gauge-field configurations.  For all combinations
of two tetraquark operators in the subsets retained, we never found that two additional 
levels were extracted.  Next, we chose a single tetraquark operator that we viewed
worked the best to extract the FV spectrum in each of the isotriplet and isodoublet channels.
The flavor structures of the chosen best tetraquark operators are
$\overline{s}u\overline{s}s$ and $\overline{u}u\overline{d}u$ in the isodoublet and
isotriplet channels, respectively.  Complete details of these operators are available
upon request.
In each channel, high statistics estimates of the correlation matrix elements were finally 
obtained involving all of the meson and meson-meson operators, as well as the selected best 
tetraquark operator.

\begin{figure}
\begin{center}
 \includegraphics[scale=0.40]{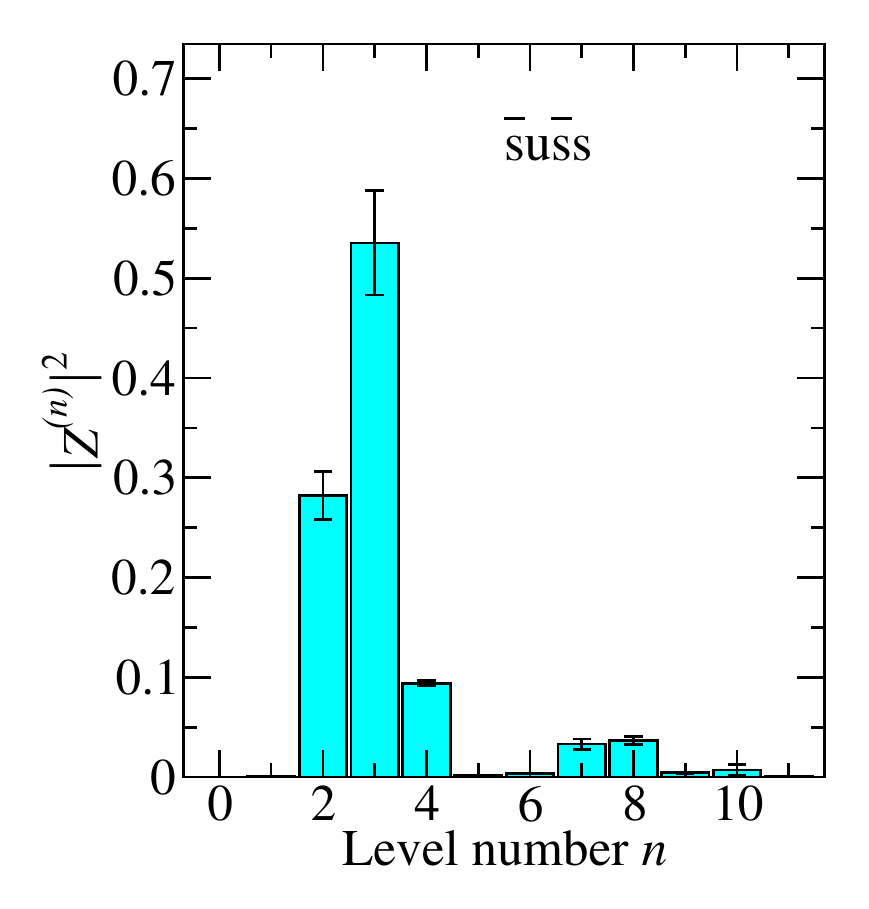}\qquad
 \includegraphics[scale=0.40]{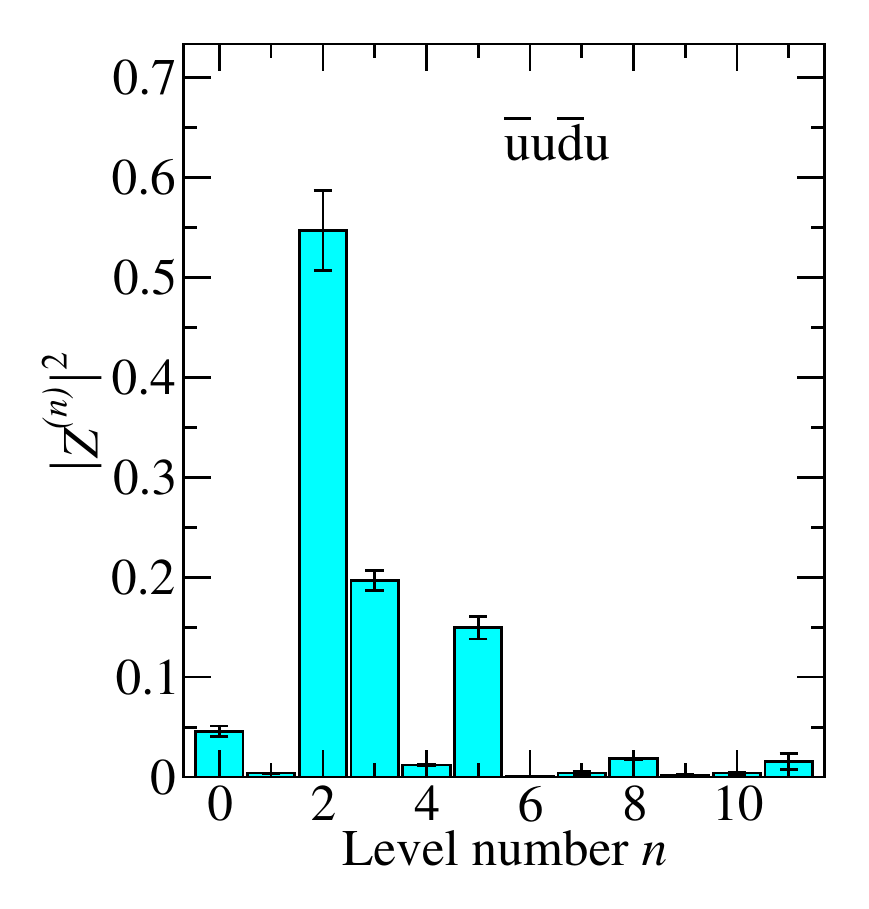}
\end{center}
\caption{Overlap factors for the single-site isodoublet strange $\overline{s}u\overline{s}s$ 
   tetraquark operator (left) and single-site isosinglet nonstrange $\overline{u}u\overline{d}u$ 
   tetraquark operator (right). 
   \label{fig:tqzfactors}} 
\end{figure}

In the same way that we analyzed the $\Delta$ operator in the isoquartet nonstrange $N\pi$ 
scattering channel, we analyzed the effect of including the tetraquark operators on our ability 
to extract the FV spectrum for these mesonic scattering channels.  The results are shown
in Fig.~\ref{fig:spectrums}.  The extracted spectra with and without the tetraquark operators
are shown in the center plot of Fig.~\ref{fig:spectrums} for the isodoublet strange channel, and
in the right plot for the isotriplet nonstrange channel.  In both cases, the presence of
an additional level is clearly observed when including the tetraquark level.  
Overlaps factors for the tetraquark operators are shown in Fig.~\ref{fig:tqzfactors}.
Note that several
quark-antiquark operators are used in each channel, and a large number of meson-meson
operators are used.  It seems that it is not possible to produce this additional level 
using more and more single and two-meson operators.  Also, each additional level lies well
below the thresholds for three-meson and four-meson energies, so it seems very unlikely that
three-meson and four-meson operators could produce the additional levels.  

\begin{figure}
 \centering
 \includegraphics[width=0.6\linewidth]{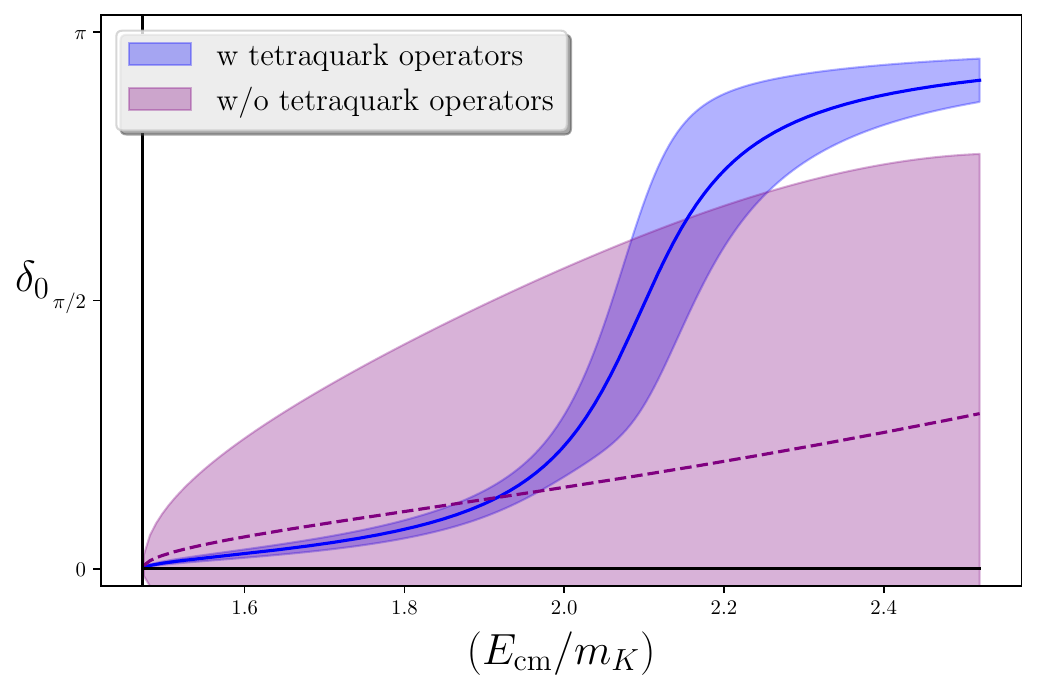}
 \caption{Preliminary results for the $\kappa$ resonance $s$-wave phase shift against the
 center-of-mass energy over the kaon mass, $m_K$, calculated using the operator set with and 
 without the tetraquark operator. 
 \label{fig:kappa_phase}}
\end{figure}

Again, to reliably obtain information on the $K_0^*(700)$ and $a_0(980)$ resonances, 
FV channels involving a variety of nonzero total momenta are needed.  However, it is still a useful
exercise to perform a L\"uscher analysis using just the zero momentum channels.
For this qualitative analysis, the following simple parametrizations in terms of the
Mandelstam variable $s$ are used:
\begin{center}
\begin{minipage}[t]{0.45\textwidth}
 \vspace*{-3em}
 \begin{equation}
     \label{eq:kmat1}
      \Tilde{K}^{-1}(s) = \textup{diag}( -A_1(s-s_0), 1/a_1),
 \end{equation}
\end{minipage}
\begin{minipage}[t]{0.45\textwidth}
 \vspace*{-2em}
 \begin{equation}
     \label{eq:kmat2}
     \Tilde{K}^{-1}(s) = \textup{diag}( 1/a_0, 1/a_1),
 \end{equation}
\end{minipage}\\
\vspace*{-1em}
\end{center}
where $\widetilde{K}$ is related to the $K$-matrix by the removal of simple threshold
factors.  Preliminary results for the $K_0^\ast(700)$ fits using the 5 lowest levels from 
the spectrum without the tetraquark operators and then the lowest 6 levels with the tetraquark 
operators are presented in Table~\ref{tab:fit_res}, and the resulting
phase-shift determinations are shown in Fig.~\ref{fig:kappa_phase}. 
It is clear from this plot that with the tetraquark operator, the characteristic behavior
of a resonance is seen, and without the tetraquark operator, there appears to be no resonance 
whatsoever.  Similarly, no resonances were found in 
Refs.~\cite{Brett:2018jqw,Wilson:2019wfr} which used no tetraquark operators.
Looking at
the fit qualities, the fit that includes the tetraquark operator is preferred. 

\begin{table}
 \caption{\label{tab:fit_res}Preliminary results for the scattering amplitude fit parameters 
  for the $\kappa$ and $a_0$ resonance channels. $N_{\textup{TQO}}$ is the number of tetraquark
  operators included in the correlation matrix of the analysis. The column labelled `equation' 
  indicates the 
  equation used for the fit. $m_K$ is the mass of the kaon, and $A_1$, $s_0$, $a_0$, and $a_1$ 
  are the fit parameters shown in Eq.~(\ref{eq:kmat1}) or (\ref{eq:kmat2}). The 
  $\chi^2/\textup{d.o.f.}$ indicates the fit quality.}
 \centering
 \begin{tabular}{c c c c c c c c}
     channel & $N_{\textup{TQO}}$ & equation & $A_1$ & $s_0$ & $m_Ka_0$ & $m_Ka_1$ & $\chi^2/\textup{d.o.f.}$ \\
     \hline
     $\kappa$ & 1 & \ref{eq:kmat1} & 1.7(7) & 4.42(24) & & 6.1(3.6) & $3.08/(6 - 3)$  \\
     $\kappa$ & 0 & \ref{eq:kmat1} & -0.20(18) & 10(5) & & -0.3(5) & $5.81/(5 - 3)$   \\
     $a_0$ & 1 & \ref{eq:kmat2} &  &  & 2.1(5) & -1(3) & $1.54/(3 - 2)$  \\
      $a_0$ & 0 & \ref{eq:kmat2} &  &  & -0.1(1.9) & -0.1(1.9) & $2.67/(3 - 2)$  
 \end{tabular}
\end{table}

The preliminary results for the $a_0(980)$ channel were not as constrained due to the low 
quality of the FV fits and only 3 levels were used for either fit. With the tetraquark 
operator, the fit produces a virtual bound state, and without the tetraquark operator, no 
virtual bound state is produced. The results of these fits are in Table~\ref{tab:fit_res}, 
and based on the fit quality, the fit that includes the tetraquark operator is preferred 
here as well. 

\section{Conclusion}
A second look at the $N\pi$ scattering channel verifies the well-known importance of using 
a three-quark $\Delta$ operator, in addition to the $N\pi$ operators, in order
to reliably extract the finite-volume spectra in any study of the $\Delta(1232)$
resonance.  Similarly, our findings point to the surprising fact that any study of the 
$K_0^\ast(700)$ and $a_0(980)$ resonances using the L\"uscher formalism should take 
tetraquark operators into account to reliably obtain the needed finite-volume spectra.

\acknowledgments
Calculations for the results presented here were performed on the HPC
clusters ``HIMster II'' at the Helmholtz-Institut Mainz, ``Mogon II''
at JGU Mainz, and ``Frontera'' at the Texas Advanced Computing Center (TACC).  
The computations were performed using the \texttt{chroma\_laph} and 
\texttt{last\_laph} software suites.
\texttt{chroma\_laph} uses the USQCD \texttt{chroma}~\cite{Edwards:2004sx} library and 
the \texttt{QDP++} library.
The contractions were optimized with \texttt{contraction\_optimizer}~\cite{contraction_optimizer}.
The computations were managed with \texttt{METAQ}~\cite{Berkowitz:2017vcp,Berkowitz:2017xna}.
The correlation function analysis was performed with \texttt{chimera} and \texttt{SigMonD}.
We are grateful to our colleagues within the CLS initiative for sharing ensembles.

This work was supported in part by the U.S.~NSF under awards PHY-1913158 and 
PHY-2209167 (CJM, SS, DD), the Faculty Early Career Development Program (CAREER) under 
award PHY-2047185 (AN), the U.S.~Department of Energy, 
Office of Science, Office of Nuclear Physics, under grant contract numbers DE-SC0011090 
and DE-SC0021006 (FRL), DE-SC0012704 (ADH), DE-AC02-05CH11231 (AWL) and within the 
framework of Scientific Discovery through Advanced Computing (SciDAC) award ``Fundamental 
Nuclear Physics at the Exascale and Beyond'' (ADH), and the Mauricio and 
Carlota Botton Fellowship (FRL). 

\bibliographystyle{JHEP}
\bibliography{references}

\end{document}